\documentclass[12pt]{article}
\usepackage{amsmath}
\usepackage{amssymb}
\usepackage{amsthm}
\usepackage{graphicx,psfrag,epsf}
\usepackage{psfrag,epsf}
\usepackage{enumitem}
\usepackage{natbib}
\usepackage{url} 
\usepackage[titletoc,title]{appendix}
\usepackage{array}
\usepackage{bm}
\usepackage[dvipsnames]{xcolor}
\usepackage{pifont}
\usepackage{color}
\usepackage{graphicx}
\usepackage{epstopdf}
\usepackage[letterpaper, portrait, margin=1in]{geometry}
\usepackage{lipsum}
\usepackage{listings}
\usepackage{fancyref}
\usepackage{float}
\usepackage[font=footnotesize,labelfont=bf]{caption}
\usepackage{color}
\usepackage{xr}
\usepackage{mathtools}

 \usepackage{caption}
 \usepackage{subcaption}
 
\usepackage{color}

\newtheorem{theorem}{Theorem}[section]

\DeclarePairedDelimiter{\ceil}{\lceil}{\rceil}

\theoremstyle{remark}
\newtheorem*{remark}{Remark}

\begin{document}

\def\spacingset#1{\renewcommand{\baselinestretch}%
{#1}\small\normalsize} \spacingset{1}

  \title{Note on the Delta Method for Finite Population Inference with Applications to Causal Inference\thanks{\noindent{\textit{Email: } \texttt{nicole.pashley@rutgers.edu}. The author would like to thank Espen Bernton for insightful discussions and comments, especially with regards to the Skorokhod representation step of the proof.
The author would also like to thank Sanqian Zhang, Luke Miratrix, and the Miratrix C.A.R.E.S. Lab for helpful comments on the work.
Nicole Pashley was supported by the National Science Foundation Graduate Research Fellowship under Grant No. DGE1745303 while working on this project.
Any opinion, findings, and conclusions or recommendations expressed in this material are those of the author and do not necessarily reflect the views of the National Science Foundation.
}}}
    \author{Nicole E. Pashley\\Department of Statistics, Rutgers University}
 \maketitle

\bigskip
\begin{abstract}
This work derives a finite population delta method.
The delta method creates more general inference results when coupled with central limit theorem results for the finite population.
This opens up a range of new estimators for which we can find finite population asymptotic properties.
We focus on the use of this method to derive asymptotic distributional results and variance expressions for causal estimators. 
We illustrate the use of the method by obtaining a finite population asymptotic distribution for a causal ratio estimator.
\end{abstract}

\noindent%
\vfill

\spacingset{1.45}

\section{Introduction}\label{sec:intro}
If you are a statistician, or even if you have simply taken an introductory probability course, you are likely well acquainted with the delta method.
Despite its relatively mysterious beginnings \citep{ver2012invented}, the delta method is a widely used tool for obtaining asymptotic distributions and variances.
In particular, the method allows us to obtain asymptotic distributions based on differentiable functions of our estimators.
The ability to work with functions of our estimators opens up a world of opportunity in terms of what we can estimate.
And why should we limit ourselves to linear estimators?
Yet, presentations of the delta method invariably, to our knowledge, work in a setting with some imagined infinite super population from which units are sampled.
However, there are many settings in which we are instead interested in understanding the behavior of finite-population inference as the population size grows.

One such setting is finite-population inference under the potential outcome framework \citep{neyman_1923, rubin_1974} for causal inference.
In this setting, inference is made only with respect to units in the study, these units and their potential outcomes are considered fixed, and randomness comes only from random assignment of units to treatment.
Fisher and Neyman were leaders in developing inference methodology for this setting \citep{Fisher1992, neyman_1923} and, following them, much of the causal inference literature has focused on estimating additive treatment effects or the difference in means.
\citet{li2017general} laid out general central limit theorems for these types of causal estimators in the finite population setting.
They also explored other causal estimators such as those for multiple treatments and with regression adjustment.
In this paper, we aim to build upon the work of \citet{li2017general} by giving conditions for a finite-population version of the delta method that can be used with their central limit theorems.

We start by giving the results for the univariate case in Section~\ref{sec:delta_uni}.
Then we give results for the multivariate case in Section~\ref{sec:delta_multi}.
In this section we also discuss applications to causal inference.
Section~\ref{sec:disc} concludes.
Proofs can be found in the supplementary material.
The theorems are intentionally kept general, with examples illustrating applications in the causal inference setting.

\section{Univariate delta method}\label{sec:delta_uni}

\subsection{Simple setting}\label{sec:uni_simple}
We start by giving a simple and standard setting for finite population inference for means.
Let us have a fixed, finite population of $N$ units, with associated values $\{A_{N, 1}, \cdots, A_{N, N}\}$.
As typical in the finite population literature \citep[see][]{aronow2013class, erich1975nonparametrics, erich1999largesample, li2017general, scott1981asymptotic}, assume that this finite population is embedded in a sequence of growing finite populations.
From the finite population, we draw a random sample of $n = \ceil*{pN}$ units, for some fixed $0 < p <1$.
By fixing $p$, $n \to \infty$ and $N - n \to \infty$ as $N \to \infty$.
We are interested in estimating the finite population mean, $A_N$, using the sample mean, $a_N$.
Let $Z_i$ be the indicator for inclusion of unit $i$ in the sample, i.e. $Z_i = 1$ if unit $i$ is in the sample and $Z_i=0$ if unit $i$ is not in the sample.
Then, because each unit $i$ in the population has associated value $A_{N, i}$, the population mean would be $A_N =N^{-1}\sum_{i=1}^NA_{N, i}$
and the sample mean would be
$a_N = n^{-1}\sum_{i=1}^NZ_iA_{N, i}$.
Because the finite population outcomes are fixed, $A_N$ is not random.
Subscripting all terms by $N$ is used to clarify that we take limits with respect to the growing sequence of finite populations, even though $a_N$  only uses a sample of those $N$ units.

\subsection{General setting}\label{sec:uni_delt}
Assume that we have a fixed finite population of $N$ units embedded in a sequence of growing finite populations.
We are interested in estimating some finite-population value $A_N \in \mathbb{R}$, which is a fixed function of the outcomes, or other values, associated with the $N$ units.
Again, $A_N$ is not random but rather fixed for the finite population of size $N$.
Let $a_N$ be some random estimator for $A_N$.
For instance, $A_N$ could be the finite-population average treatment effect and $a_N$ could be an estimate of the average treatment effect based on random assignment of units to treatment and control.
We have a function $g: \mathbb{R} \to \mathbb{R}$ and wish to make inference for $g(a_N) - g(A_N)$.
Throughout, we assume limits are taken as $N \to \infty$.

\begin{theorem}\label{theor:uni}
 Assume that we have the result that, as $N \to \infty$,
\[a_N- A_N \xrightarrow[]{p} 0\]
and
\begin{align}
\frac{a_N-A_N}{\sqrt{\text{Var}(a_N)}} = \frac{\sqrt{N}(a_N-A_N)}{\sqrt{N\text{Var}(a_N)}} \xrightarrow[]{d} \text{N}(0,1). \label{eq:norm_con}
\end{align}
Assume that there is sequence of subsets of $\mathbb{R}$, $\{R_N(\epsilon)\}$, such that for almost all $N$, $A_N \in R_N(\epsilon)$ and $P(a_N \in R_N(\epsilon)) \geq 1-\epsilon$.
Let $g: \mathbb{R} \to \mathbb{R}$ be a differentiable function that is uniformly continuous with respect to $\{R_N(\epsilon)\}$, for every $\epsilon >0$, with $g'$ also uniformly continuous with respect to $\{R_N(\epsilon)\}$, for every $\epsilon >0$.
Also assume that $g'(A_N)$ is bounded away from zero such that $\liminf_{N \to \infty} |g'(A_N)| \geq c > 0$.
Then we have the result
\begin{align}
\frac{g(a_N) - g(A_N)}{g'(A_N)\sqrt{\text{Var}(a_N)}} \xrightarrow[]{d} \text{N}(0,1). \label{eq:uni_result}
\end{align}
\end{theorem}

\begin{proof}
Following the usual delta method proof \citep[e.g.,][]{gut2005probability}, we have from Taylor's Theorem and the Mean Value Theorem that
$g(a_N) = g(A_N) + g'(b_N)(a_N-A_N)$
for some random variable $b_N$ which is point-wise between $A_N$ and $a_N$.
We must have $b_N - A_N$ goes to 0 in probability because $|b_N - A_N| \leq |a_N- A_N|$ and $|a_N- A_N|$ goes to 0 in probability, by assumption.
By rearranging the Taylor's Theorem result,
\[\frac{g(a_N) - g(A_N)}{g'(b_N)\sqrt{\text{Var}(a_N)}} = \frac{a_N - A_N}{\sqrt{\text{Var}(a_N)}} \xrightarrow[]{d} \text{N}(0,1).\]

Now we want to change the $b_N$ in the denominator to $A_N$.
This step is necessary in the finite-population case because we standardize by the finite population variance and do not necessarily assume a limiting value on $A_N$.
The difference is
\begin{align*}
\frac{g(a_N) - g(A_N)}{g'(b_N)\sqrt{\text{Var}(a_N)}}  - \frac{g(a_N) - g(A_N)}{g'(A_N)\sqrt{\text{Var}(a_N)}} 
=\frac{g'(A_N) - g'(b_N)}{g'(A_N)} \frac{g(a_N) - g(A_N)}{g'(b_N)\sqrt{\text{Var}(a_N)}}.
\end{align*}
For these terms to be defined we need to assume that $g'(A_N) \neq 0$ and $g'(b_N) \neq 0$ (or the same for $g'(a_N)$).
We have shown that $(g(a_N) - g(A_N))/(g'(b_N)\sqrt{\text{Var}(a_N)})$ is asymptotically Normally distributed.
Using the fact that $b_N- A_N$ goes to 0 in probability and the uniform continuity of the derivative, we can apply Theorem 2 of \cite{mann1943stochastic} and find that 
$g'(b_N) - g'(A_N)$ goes to 0 in probability.
We assumed that $g'(A_N)$ is bounded away from 0 as $N \to \infty$\footnote{We only need require that $g'(A_N)$ cannot converge to 0 faster than $g'(A_N) - g'(b)$ converges to 0 in probability.}, so the entire difference must go to 0 in probability.
This gives the desired result.
\end{proof}
We can use Theorem 1 of \citet{li2017general} to get the normality condition of Equation~\ref{eq:norm_con} for sample means.

\begin{remark}
The condition of uniform continuity of $g$ and $g'$ can be satisfied by assuming that $A_n$ and $a_n$ belong to a compact set, such as a closed interval of $\mathbb{R}$ and that both $g$ and $g'$ are continuous over that compact set.
\end{remark}

\begin{remark}
In the case that $g'(A_N) = 0$, we may be able to use a higher order delta method.
\end{remark}

\begin{remark} Typical delta method approaches for super-population settings assume a limiting value for $a_N$, say $\mu_a$ and only need to assume that $g'(\mu_a) \neq 0$.
Further, Theorem~\ref{theor:uni} standardizes by the finite-population variance.
This is different than the standard formulation of the delta method with a limiting value of the variance included in the asymptotic normal distribution.
Theorem~\ref{theor:uni} thus has stronger conditions on continuity than the standard delta method and bounds on the derivative away from zero, but does not require limiting values for means or variances.
\end{remark}

\subsection{Example: Squared estimator}\label{sec:ex_squared}
Assume that we have a random sample of $n = \ceil*{pN}$ units from a population of $N$ units such that $1 \leq n \leq N-1$, with the assumption that $p$ remains fixed as $N \to \infty$.
Each unit has associated outcome $B_1 \geq Y_{N,i} \geq B_2 >0$, where $B_1$ and $B_2$ are finite, positive bounds on the outcome.
Let $Z_i = 1$ if unit $i$ is included in the sample.
Let $\bar{Y}_N$ be the population mean outcome (i.e., $\bar{Y}_N = \sum_{i=1}^N Y_{N,i}/N$) and $\bar{y}_N$ be the observed or estimated mean outcome (i.e., $\bar{y}_N = \sum_{i=1}^N Z_iY_{N,i}/n$).
We are interested in finding the finite population asymptotic distributional result for $\bar{y}_N^2$.
Following notation from \citet{li2017general}, let
\[m_N = \text{max}_{1 \leq i \leq N}\left[Y_{N,i} - \bar{Y}_N\right]^2\]
and
\[v_N = \frac{1}{N-1}\sum_{i=1}^N\left[Y_{N,i} - \bar{Y}_N\right]^2 .\]
We have from Theorem 1 of \citet{li2017general} that if as $N \to \infty$
\[\frac{1}{\text{min}(n, N-n)}\frac{m_N}{v_N} \to 0\]
 then
\[\frac{\bar{y}_N-\bar{Y}_N}{\sqrt{\text{Var}(\bar{y}_N)}} \xrightarrow[]{d} \text{N}(0, 1).\]
We also assume that $n\text{Var}(\bar{y}_N)$ (defined below) has a finite limiting value, which ensures $\bar{y}_N-\bar{Y}_N$ goes to 0 in probability,
which can be proved using the previous distributional result or by Markov's inequality.

Define $g(x) = x^2$.
Then $g'(x) = 2x$.
Because $g(x)$ and $g'(x)$ are continuous for all finite values of $x$ and the outcomes (and therefore the means) are constrained to a closed, bounded interval which does not contain zero, the uniform continuity requirement is satisfied as well as the requirement that $\liminf_{N \to \infty} |g'(A_N)| \geq 2B_2 > 0$.

It is well known and can be found in, for instance, \citet{li2017general} that
\[\text{Var}(\bar{y}_N) = \left(\frac{1}{n} - \frac{1}{N}\right)v_N.\]
Then we can easily apply Theorem~\ref{theor:uni} to get
\begin{align*}
\frac{\bar{y}_N^2 - \bar{Y}_N^2}{2\bar{Y}_N\sqrt{\text{Var}(\bar{y}_N)}} \xrightarrow[]{d} \text{N}(0,1).
\end{align*}

\section{Delta method: Multivariate}\label{sec:delta_multi}

We now give a more general version of the delta method when we have a vector of outcomes.
That is, now $\bm{A}_N, \bm{a}_N \in \mathbb{R}^K$ with $\bm{A}_N$ still a fixed quantity based on the $N$ units in the finite population and $\bm{a}_N$ still a random estimator for $\bm{A}_N$.
For instance, as in Section 3 of \citet{li2017general}, $\bm{a}_N$ could be a vector of observed averages of potential outcomes under different treatments (or a linear combination there of) and $\bm{A}_N$ could be the corresponding true finite population averages of potential outcomes.
We again assume that we have a normality result for $\bm{a}_N-\bm{A}_N$.
Let the $k$th component $\bm{a}_N$ and $\bm{A}_N$ be denoted $\bm{a}_{N[k]}$ and $\bm{A}_{N[k]}$, respectively.
We are interested in finding a similar result for some function of our estimator, $g: \mathbb{R}^K \to \mathbb{R}$, with conditions on $g$ laid out in Theorem~\ref{theor:main}.
Let $\nabla g(\bm{b})$ be the vector of partial derivatives evaluated at $\bm{b}$.
Define
\begin{align*}
\bm{V}_N &= \begin{pmatrix}
\sqrt{\text{Var}(\bm{a}_{N[1]})} & 0 & \cdots & 0\\
0 & \sqrt{\text{Var}(\bm{a}_{N[2]})}  & \cdots & 0\\
\vdots & \vdots & \ddots & \vdots\\
0 & 0 & \cdots & \sqrt{\text{Var}(\bm{a}_{N[K]})}
\end{pmatrix}.
\end{align*}

\begin{theorem}\label{theor:main}
 Assume that we have the result that, as $N \to \infty$,
\[\bm{a}_N - \bm{A}_N \xrightarrow[]{p} \bm{0}\]
and
\[\left(\frac{\bm{a}_{N[1]}-\bm{A}_{N[1]}}{\sqrt{\text{Var}(\bm{a}_{N[1]})}}, \dots, \frac{\bm{a}_{N[K]}-\bm{A}_{N[K]}}{\sqrt{\text{Var}(\bm{a}_{N[K]})}}\right)^T \xrightarrow[]{d} \text{N}(\bm{0},\bm{\Sigma}),\]
where $\bm{\Sigma}$ is the limit of the correlation matrix. 
Assume that there is sequence of compact subsets of $\mathbb{R}^K$, $\{R_N(\epsilon)\}$, such that for almost all $N$, $\bm{A}_N \in R_N(\epsilon)$ and $P(\bm{a}_N \in R_N(\epsilon)) \geq 1-\epsilon$.
Let $g:  \mathbb{R}^K \to \mathbb{R}$ be a differentiable function that is uniformly continuous with respect to $\{R_N(\epsilon)\}$, for every $\epsilon >0$, with all of its first order partial derivatives also uniformly continuous with respect to $\{R_N(\epsilon)\}$, for every $\epsilon >0$.
Also assume that 
$ \left(\nabla g(\bm{b}_N)-\nabla g(\bm{A}_N)\right)^T \bm{V}_N/\sqrt{\left(\nabla g(\bm{A}_N)\right)^T \bm{V}_N\bm{\Sigma}\bm{V}_N\nabla g(\bm{A}_N)} \overset{p}{\to} 0$, for which it is sufficient that\\ $\sqrt{\text{Var}(\bm{a}_{N[k]})}/\sqrt{\left(\nabla g(\bm{A}_N)\right)^T \bm{V}_N\bm{\Sigma}\bm{V}_N\nabla g(\bm{A}_N)} = O(1)$ for any $k$ such that $\nabla g(\bm{x})$ is not identically 0 for $\bm{x} \in \{R_N(\epsilon)\}$.

Then we have the results
 \[\frac{g(\bm{a}_N) - g(\bm{A}_N)}{\sqrt{\left(\nabla g(\bm{A}_N)\right)^T \bm{V}_N\bm{\Sigma}\bm{V}_N\nabla g(\bm{A}_N)}} \xrightarrow[]{d} \text{N}(0,1) \quad \text{ and} \quad \frac{g(\bm{a}_N) - g(\bm{A}_N)}{\sqrt{\left(\nabla g(\bm{A}_N)\right)^T \bm{V}_N\bm{\Sigma}_N\bm{V}_N\nabla g(\bm{A}_N)}} \xrightarrow[]{d} \text{N}(0,1),\]
 where $\bm{\Sigma}_N$ is the finite-population correlation matrix and so $\bm{V}_N\bm{\Sigma}_N\bm{V}_N$ is the covariance matrix of $\bm{a}_N - \bm{A}_N$.
\end{theorem}

\begin{remark}
Having $\sqrt{\text{Var}(\bm{a}_{N[k]})}/\sqrt{\left(\nabla g(\bm{A}_N)\right)^T \bm{V}_N\bm{\Sigma}\bm{V}_N\nabla g(\bm{A}_N)} = O(1)$ translates in the univariate case to $1/g'(A_N) = O(1)$, which implies $g'(A_N)$ cannot converge to 0.
If we assume $\sqrt{N}\bm{V}_N$ has limiting value $\bm{V}$ and $g(\bm{A}_N)$ has limiting value $g(\bm{A})$, we can write the asymptotic variance of $\sqrt{N}[g(\bm{a}_N) - g(\bm{A}_N)]$ as $\left(\nabla g(\bm{A})\right)^T \bm{V}\bm{\Sigma}\bm{V}\nabla g(\bm{A})$. Thus, this requirement regards the degeneracy of the asymptotic distribution. See Supplementary Material for an example where degeneracy can occur.
\end{remark}

The remarks from the univariate case extend directly to the multivariate case.
See Supplementary Material A for derivations.
The derivation is more complicated for the finite-population setting than the standard super-population setting because the lack of limiting values on means and variances implies we must standardize the expression.
The value we standardize by is a function of $g(\bm{A}_N)$ and $\bm{V}_N$, which are sequences of values rather than fixed constants.


\subsection{General form of causal estimator variance}\label{sec:causal_var}
In this section we look at the classic causal inference set up with two treatment groups.
We derive the general form of the variance for a function of the observed means of potential outcomes under treatment and control.
Let us have $N$ units in the finite population with $n_1=pN (>1)$ units assigned to treatment and $n_0=(1-p)N (>1)$ units assigned to control, with the assumption that $p$ remains fixed as $N \to \infty$.
Let $Z_i = 1$ if unit $i$ is assigned to treatment and $Z_i = 0$ if unit $i$ is assigned to control.
The potential outcome for unit $i$ under treatment is $Y_{N,i}(1)$ and under control is $Y_{N,i}(0)$.
Let $\bar{Y}_N(z)$ be the population mean potential outcome under treatment $z$ (i.e., $\bar{Y}_N(z) = \sum_{i=1}^N Y_{N,i}(z)/N$) and $\bar{y}_N(z)$ be the observed or estimated mean potential outcome under treatment $z$ (e.g., $\bar{y}_N(1) = \sum_{i=1}^N Z_iY_{N,i}(1)/n_1$).
We are interested in the asymptotic distribution for some function, $g: \mathbb{R}^2 \to \mathbb{R}$, of the estimated potential outcome means.
We can proceed with inference in this scenario, keeping with the randomization based framework, by utilizing the finite population delta method.

First we need to satisfy the conditions of Theorem~\ref{theor:main}.
Following notation from \citet{li2017general}, let
\[m_{N,z} = \text{max}_{1 \leq i \leq N}\left[Y_{N,i}(z) - \bar{Y}_N(z)\right]^2, \quad z \in \{0, 1\}\]
and
\[v_{N,z} = \frac{1}{N-1}\sum_{i=1}^N\left[Y_{N,i}(z) - \bar{Y}_N(z)\right]^2, \quad z \in \{0, 1\}.\]
We have from Theorem 4 of \citet{li2017general} that if as $N \to \infty$
\[\text{max}_{z \in \{0,1\}}\frac{1}{n_z}\frac{m_{N,z}}{v_{N,z}} \to 0\]\
and
the correlation matrix of $(\bar{y}_N(0), \bar{y}_N(1))$ has limiting value $\bm{\Sigma}$, then
\[\left(\frac{\bar{y}_N(0)-\bar{Y}_N(0)}{\sqrt{\text{Var}(\bar{y}_N(0))}}, \frac{\bar{y}_N(1)-\bar{Y}_N(1)}{\sqrt{\text{Var}(\bar{y}_N(1))}}\right) \xrightarrow[]{d} \text{N}(0, \bm{\Sigma}).\]
We also assume that $N\text{Var}(\bar{y}_N(0))$ and $N\text{Var}(\bar{y}_N(1))$ have finite limiting values to ensure the required convergence in probability assumption:
\[\bar{y}_N(0)-\bar{Y}_N(0) \xrightarrow[]{p} 0 \quad \text{and} \quad \bar{y}_N(1)-\bar{Y}_N(1) \xrightarrow[]{p} 0.\]
With appropriate assumptions on the domain of the means and continuity of  $g$ and its partial derivatives, we can invoke Theorem~\ref{theor:main}.

We have
\begin{align*}
\text{Var}\left(\bar{y}_N(0)\right) &= \frac{p}{(1-p)N(N-1)}\sum_{i=1}^N\left(Y_{N,i}(0)-\bar{Y}_N(0)\right)^2= \left(\frac{1}{n_0} - \frac{1}{N}\right)v_{N,0}\\
\text{Var}\left(\bar{y}_N(1)\right) &= \frac{1-p}{pN(N-1)}\sum_{i=1}^N\left(Y_{N,i}(1)-\bar{Y}_N(1)\right)^2 = \left(\frac{1}{n_1} - \frac{1}{N}\right)v_{N,1}\\
\text{Cov}\left(\bar{y}_N(1), \bar{y}_N(0)\right) &= -\frac{1}{N(N-1)}\sum_{i=1}^N\left(Y_{N,i}(1)-\bar{Y}_N(1)\right)\left(Y_{N,i}(0)-\bar{Y}_N(0)\right).
\end{align*}


$\nabla g\left(\bar{Y}_N(0), \bar{Y}_N(1)\right)$ is a vector of length two, with the first entry, which we denote $\nabla g_{[1]}$, corresponding to the partial derivative with respect to the control mean and the second entry, $\nabla g_{[2]}$, corresponding to the partial derivative with respect to the treatment mean.
For the covariance term,
\[\bm{V}_N\bm{\Sigma}_N\bm{V}_N = \begin{pmatrix}
\text{Var}\left(\bar{y}_N(0)\right) & \text{Cov}\left(\bar{y}_N(1), \bar{y}_N(0)\right)\\
\text{Cov}\left(\bar{y}_N(1), \bar{y}_N(0)\right)& \text{Var}\left(\bar{y}_N(1)\right)
\end{pmatrix}.\]

So then
\begin{align*}
&\left(\nabla g\left(\bar{Y}_N(0), \bar{Y}_N(1)\right)\right)^T\bm{V}_N\bm{\Sigma}\bm{V}_N\nabla g\left(\bar{Y}_N(0), \bar{Y}_N(1)\right)\\
&= \nabla g_{[1]}^2\text{Var}\left(\bar{y}_N(0)\right) + 2\nabla g_{[1]}\nabla g_{[2]}\text{Cov}\left(\bar{y}_N(1), \bar{y}_N(0)\right) + \nabla g_{[2]}^2\text{Var}\left(\bar{y}_N(1)\right)\\
&= \nabla g_{[1]}^2\frac{v_{N,0}}{n_0} + \nabla g_{[2]}^2\frac{v_{N,1}}{n_1} - \frac{1}{N(N-1)}\sum_{i=1}^N\left[\nabla g_{[2]}Y_{N,i}(1) + \nabla g_{[1]}Y_{N,i}(0) - \left(\nabla g_{[2]}\bar{Y}_{N}(1) + \nabla g_{[1]}\bar{Y}_{N}(0)\right)\right]^2.
\end{align*}

We can use the typical Neyman style variance estimators to estimate $v_{N,1}$ and $v_{N,0}$.
Hence we can obtain a conservative estimator for the variance term by excluding the negative term. 
That is, for $v_{N,z}$ we can use estimator
\[\hat{v}_{N, z}  = \frac{1}{n_z-1}\sum_{i:Z_i = z}\left(Y_{N,i}(z)-\bar{y}_N(z)\right)^2.\]

Further, this variance is equivalent to the standard Neyman, randomization variance for an experiment with potential outcomes $\tilde{Y}_{N,i}(1) =  \nabla g_{[2]}Y_{N,i}(1)$ and $\tilde{Y}_{N,i}(0) = \nabla g_{[1]}Y_{N,i}(0)$.

\subsection{Example: Ratio Estimator}\label{sec:ex_ratio}
We now give an example, using the causal inference set up and notations introduced in the previous section. 
To apply Theorem~\ref{theor:main}, we make the same assumptions to get the normality and convergence results given in the prior section.
Most of the causal inference literature tries to estimate the average treatment effect defined as $\tau = \bar{Y}_N(1) - \bar{Y}_N(0)$.
However, a multiplicative effect, $\tau  = \bar{Y}_N(1)/\bar{Y}_N(0)$, may also be of interest\footnote{Note that $\bar{Y}_N(1)/\bar{Y}_N(0)$ is not in general equivalent to the average of $Y_{N,i}(1)/Y_{N,i}(0)$.} for outcomes such that $B_1 \geq Y_{N,i}(z) \geq B_2 >0$ for $z \in\{0,1\}$, where $B_1$ and $B_2$ are finite, positive bounds on the outcome.
Set function $g(x, w) = x/w$.
Then
\[\left(\nabla g(x, w)\right)^T = \left(-\frac{x}{w^2}, \frac{1}{w} \right).\]
Given the bounds on $Y_{N,i}(z)$ for $z \in\{0,1\}$, the continuity requirements are satisfied for the original function and the partial derivatives.
We further assume that\\ $\sqrt{v_{N,1}} /\sqrt{\frac{1}{\bar{Y}_N(0)^2}\left(\frac{v_{N,1}}{p} + \frac{\bar{Y}_N(1)^2}{\bar{Y}_N(0)^2}\frac{v_{N,0}}{1-p}\right) -\frac{1}{N-1}\sum_{i=1}^N\frac{Y_{N,i}(0)^2}{\bar{Y}_N(0)^2}\left(\frac{Y_{N,i}(1)}{\bar{Y}_{N,i}(0)}-\frac{\bar{Y}_N(1)}{\bar{Y}_N(0)}\right)^2} = O(1)$, and \\$\sqrt{v_{N,0} }/\sqrt{\frac{1}{\bar{Y}_N(0)^2}\left(\frac{v_{N,1}}{p} + \frac{\bar{Y}_N(1)^2}{\bar{Y}_N(0)^2}\frac{v_{N,0}}{1-p}\right) -\frac{1}{N-1}\sum_{i=1}^N\frac{Y_{N,i}(0)^2}{\bar{Y}_N(0)^2}\left(\frac{Y_{N,i}(1)}{\bar{Y}_{N,i}(0)}-\frac{\bar{Y}_N(1)}{\bar{Y}_N(0)}\right)^2}= O(1)$.

We have under Theorem~\ref{theor:main} and using the simplified variance expressions derived in the previous section,
\begin{align*}
& \frac{\bar{y}_N(1)/\bar{y}_N(0)-\bar{Y}_N(1)/\bar{Y}_N(0)}{\sqrt{\left(\nabla g(\bar{Y}_N(0), \bar{Y}_N(1))\right)^T \bm{V}_N\bm{\Sigma}_N\bm{V}_N\nabla g(\bar{Y}_N(0), \bar{Y}_N(1))}}\\
&= \frac{\bar{y}_N(1)/\bar{y}(0)-\bar{Y}_N(1)/\bar{Y}_N(0)}{\sqrt{\frac{1}{\bar{Y}_N(0)^2}\left(\frac{v_{N,1}}{n_1} + \frac{\bar{Y}_N(1)^2}{\bar{Y}_N(0)^2}\frac{v_{N,0}}{n_0}\right) -\frac{1}{N(N-1)}\sum_{i=1}^N\frac{Y_{N,i}(0)^2}{\bar{Y}_N(0)^2}\left(\frac{Y_{N,i}(1)}{\bar{Y}_{N,i}(0)}-\frac{\bar{Y}_N(1)}{\bar{Y}_N(0)}\right)^2}}\\
&\xrightarrow[]{d} \text{N}(0, 1).
\end{align*}
The variance is equivalent to the  variance for an experiment with potential outcomes $\tilde{Y}_{N,i}(1) = Y_{N,i}(1)$ and $\tilde{Y}_{N,i}(0) = Y_{N,i}(0)(\bar{Y}_N(1)/\bar{Y}_N(0))$, divided by $\bar{Y}_N(0)^2$.

\section{Discussion}\label{sec:disc}
In this work we have derived a finite population version of the delta method and have applied it to obtain results for more general causal estimators by coupling it with central limit theorem results from \citet{li2017general}.
This is useful for deriving both asymptotic distributional results and variance expressions.
There are a few generalizations of this work that could be made.
We only allow $g: \mathbb{R}^K \to \mathbb{R}$ but generalizations to $g: \mathbb{R}^K \to \mathbb{R}^J$ can be made.
Additionally, as mentioned previously, issues with partial derivatives approaching zero may be resolved by implementing a higher order delta method.
Finally, the delta method is not restricted to use with the normal distribution, and so extensions to other distributions should be explored.

\bibliographystyle{apalike}
\bibliography{finite_ext_ref}{}

\begin{thebibliography}{}

\bibitem[Aronow and Middleton, 2013]{aronow2013class}
Aronow, P.~M. and Middleton, J.~A. (2013).
\newblock A class of unbiased estimators of the average treatment effect in
  randomized experiments.
\newblock {\em Journal of Causal Inference}, 1(1):135--154.

\bibitem[Fisher, 1926]{Fisher1992}
Fisher, R.~A. (1926).
\newblock The arrangement of field experiments.
\newblock {\em Journal of Ministry of Agriculture}, 33:503--513.

\bibitem[Gut, 2012]{gut2005probability}
Gut, A. (2012).
\newblock {\em Probability: {A} graduate course}.
\newblock Springer, New York.

\bibitem[Lehmann, 1975]{erich1975nonparametrics}
Lehmann, E.~L. (1975).
\newblock {\em Nonparametrics: {S}tatistical methods based on ranks}.
\newblock Holden-Day, Inc, San Francisco, CA.

\bibitem[Lehmann, 1999]{erich1999largesample}
Lehmann, E.~L. (1999).
\newblock {\em Elements of Large-Sample Theory}.
\newblock Springer, New York.

\bibitem[Li and Ding, 2017]{li2017general}
Li, X. and Ding, P. (2017).
\newblock General forms of finite population central limit theorems with
  applications to causal inference.
\newblock {\em Journal of the American Statistical Association},
  112(520):1759--1769.

\bibitem[Mann and Wald, 1943]{mann1943stochastic}
Mann, H.~B. and Wald, A. (1943).
\newblock On stochastic limit and order relationships.
\newblock {\em The Annals of Mathematical Statistics}, 14(3):217--226.

\bibitem[Rubin, 1974]{rubin_1974}
Rubin, D.~B. (1974).
\newblock Estimating causal effects of treatments in randomized and
  nonrandomized studies.
\newblock {\em Journal of Educational Psychology}, 66(5):688--701.

\bibitem[Scott and Wu, 1981]{scott1981asymptotic}
Scott, A. and Wu, C.-F. (1981).
\newblock On the asymptotic distribution of ratio and regression estimators.
\newblock {\em Journal of the American Statistical Association},
  76(373):98--102.

\bibitem[Splawa-Neyman et~al., 1990]{neyman_1923}
Splawa-Neyman, J., Dabrowska, D.~M., and Speed, T. (1923/1990).
\newblock On the application of probability theory to agricultural experiments.
  {E}ssay on principles. {S}ection 9.
\newblock {\em Statist. Sci.}, 5(4):465--472.

\bibitem[Ver~Hoef, 2012]{ver2012invented}
Ver~Hoef, J.~M. (2012).
\newblock Who invented the delta method?
\newblock {\em The American Statistician}, 66(2):124--127.

\end{thebibliography}

\begin{appendices}
\renewcommand\appendixname{Supplementary Material}

\begin{center}
{\bf \Large  Supplementary material for ``Note on the Delta Method for Finite Population Inference with Applications to Causal Inference''}

\vspace{0.5cm}
{\Large Nicole E. Pashley}
\end{center}

\section{Delta method for vector random variable}\label{append:delta_proofs_multi}
\subsection{Simple case}\label{supp:simple}
Let
\begin{align*}
\bm{V}_N &= \begin{pmatrix}
\sqrt{\text{Var}(\bm{a}_{N[1]})} & 0 & \cdots & 0\\
0 & \sqrt{\text{Var}(\bm{a}_{N[2]})}  & \cdots & 0\\
\vdots & \vdots & \ddots & \vdots\\
0 & 0 & \cdots & \sqrt{\text{Var}(\bm{a}_{N[K]})}
\end{pmatrix}
\end{align*}
so
\begin{align*}
\bm{V}_N^{-1}&= \begin{pmatrix}
\frac{1}{\sqrt{\text{Var}(\bm{a}_{N[1]})}} & 0 & \cdots & 0\\
0 & \frac{1}{\sqrt{\text{Var}(\bm{a}_{N[2]})}}  & \cdots & 0\\
\vdots & \vdots & \ddots & \vdots\\
0 & 0 & \cdots & \frac{1}{\sqrt{\text{Var}(\bm{a}_{N[K]})}}
\end{pmatrix}.
\end{align*}

Let $\bm{A}_N, \bm{a}_N \in \mathbb{R}^K$.
Assume that we have the results that
\[\bm{a}_N - \bm{A}_N \xrightarrow[]{p} 0\]
and
\[\bm{V}_N^{-1}(\bm{a}_N - \bm{A}_N) \xrightarrow[]{d} \text{N}(\bm{0},\bm{I}_K).\]
That is, we are starting with a simple case of asymptotically independent variables.

We are interested in finding a similar result for some function of our estimator, $g: \mathbb{R}^K \to \mathbb{R}$ where $g$ is a differentiable function that is uniformly continuous with respect to $\{R_N(\epsilon)\}$, for every $\epsilon >0$, with all of its first order partial derivatives also uniformly continuous with respect to $\{R_N(\epsilon)\}$, for every $\epsilon >0$.
We have

\[g(\bm{a}_N) = g(\bm{A}_N) + \left(\nabla g(\bm{b}_N)\right)^T(\bm{a}_N-\bm{A}_N)\]
for some $\bm{b}_N$ where $\bm{b}_{N[k]}$ is between $\bm{a}_{N[k]}$ and $\bm{A}_{N[k]}$ for each $k$, and where $\nabla g(\bm{b})$ is the vector of partial derivatives evaluated at $\bm{b}_N$.

We have
\[\bm{b}_N - \bm{A}_N \xrightarrow[]{p} 0\]
and uniform continuity of the partial derivatives, so we can apply Theorem 2 of \cite{mann1943stochastic} and find that
\[\nabla g(\bm{b}_N) - \nabla g(\bm{A}_N) \xrightarrow[]{p} 0.\]

Then we  have
\begin{align*}
g(\bm{a}_N) &= g(\bm{A}_N) + \left(\nabla g(\bm{b}_N)\right)^T(\bm{a}_N-\bm{A}_N)\\
g(\bm{a}_N) -g(\bm{A}_N) &=  \left(\nabla g(\bm{A}_N)\right)^T(\bm{a}_N-\bm{A}_N) + \left(\nabla g(\bm{b}_N) - \nabla g(\bm{A}_N)\right)^T(\bm{a}_N-\bm{A}_N)\\
\frac{\left(g(\bm{a}_N) -g(\bm{A}_N)\right)}{\sqrt{\left(\nabla g(\bm{A}_N)\right)^T \bm{V}_N\bm{V}_N\nabla g(\bm{A}_N)}} &=  \frac{\left(\nabla g(\bm{A}_N)\right)^T\bm{V}_N\bm{V}_N^{-1}(\bm{a}_N-\bm{A}_N)}{\sqrt{\left(\nabla g(\bm{A}_N)\right)^T \bm{V}_N\bm{V}_N\nabla g(\bm{A}_N)}}\\
&\qquad + \frac{\left(\nabla g(\bm{b}_N) - \nabla g(\bm{A}_N)\right)^T\bm{V}_N\bm{V}_N^{-1}(\bm{a}_N-\bm{A}_N)}{\sqrt{\left(\nabla g(\bm{A}_N)\right)^T \bm{V}_N\bm{V}_N\nabla g(\bm{A}_N)}}.
\end{align*}

We see that
\[\frac{\left(\nabla g(\bm{A}_N)\right)^T\bm{V}_N}{\sqrt{\left(\nabla g(\bm{A}_N)\right)^T\bm{V}_N\bm{V}_N \nabla g(\bm{A}_N)}}\]
is a unit vector that is fixed for each $N$ but does change as $N \to \infty$.
So we have that for every $N$ and $\bm{Y} \sim \text{N}(\bm{0},\bm{I}_K)$,
\[\frac{\left(\nabla g(\bm{A}_N)\right)^T\bm{V}_N}{\sqrt{\left(\nabla g(\bm{A}_N)\right)^T\bm{V}_N\bm{V}_N \nabla g(\bm{A}_N)}}\bm{Y} \sim \text{N}(0,1).\]

By Skorokhod representation, there exists a probability space on which there exist random variables $c$ and $\bm{X}$ with $\bm{c} \sim \bm{V}_N^{-1}\left(\bm{a}_N-\bm{A}_N\right)$ for all $N$ and $\bm{X} \sim \text{N}(\bm{0},\bm{I}_K)$ satisfying $\bm{c} \xrightarrow[]{a.s.} \bm{X}$.
Now we have
\begin{align*}
&\left|\frac{\left(\nabla g(\bm{A}_N)\right)^T\bm{V}_N\bm{c}}{\sqrt{\left(\nabla g(\bm{A}_N)\right)^T\bm{V}_N\bm{V}_N \nabla g(\bm{A}_N)}} - \frac{\left(\nabla g(\bm{A}_N)\right)^T\bm{V}_N\bm{X}}{\sqrt{\left(\nabla g(\bm{A}_N)\right)^T\bm{V}_N\bm{V}_N \nabla g(\bm{A}_N)}} \right|\\
&\qquad =\left| \frac{\left(\nabla g(\bm{A}_N)\right)^T\bm{V}_N}{\sqrt{\left(\nabla g(\bm{A}_N)\right)^T \bm{V}_N\bm{V}_N\nabla g(\bm{A}_N)}}\left(\bm{c} - \bm{X}\right)\right| \\
&\qquad \leq \Big|\Big|\frac{\left(\nabla g(\bm{A}_N)\right)\bm{V}_N}{\sqrt{\left(\nabla g(\bm{A}_N)\right)^T\bm{V}_N\bm{V}_N \nabla g(\bm{A}_N)}}\Big|\Big|_2||\bm{c} - \bm{X}||_2\\
&\qquad= ||\bm{c} - \bm{X}||_2\\
&\qquad \xrightarrow[]{a.s.} 0
\end{align*}
where the second line comes from the H\"{o}lder's inequality.

So we have 
\[\frac{\left(\nabla g(\bm{A}_N)\right)^T\bm{V}_N\bm{c}}{\sqrt{\left(\nabla g(\bm{A}_N)\right)^T \bm{V}_N\bm{V}_N\nabla g(\bm{A}_N)}} \xrightarrow[]{d} \text{N}(0, 1)\]
which implies that
\[\frac{\left(\nabla g(\bm{A}_N)\right)^T\bm{V}_N\bm{V}_N^{-1}(\bm{a}_N-\bm{A}_N)}{\sqrt{\left(\nabla g(\bm{A}_N)\right)^T\bm{V}_N\bm{V}_N \nabla g(\bm{A}_N)}} \xrightarrow[]{d} \text{N}(0, 1).\]

Recall we have
\begin{align*}
\frac{\left(g(\bm{a}_N) - g(\bm{A}_N)\right)}{\sqrt{\left(\nabla g(\bm{A}_N)\right)^T\bm{V}_N\bm{V}_N\nabla g(\bm{A}_N)}}& =\frac{\left(\nabla g(\bm{A}_N)\right)^T\bm{V}_N\bm{V}_N^{-1}(\bm{a}_N-\bm{A}_N)}{\sqrt{\left(\nabla g(\bm{A}_N)\right)^T \bm{V}_N\bm{V}_N\nabla g(\bm{A}_N)}}\\
&\qquad + \frac{\left(\nabla g(\bm{b}_N) - \nabla g(\bm{A}_N)\right)^T\bm{V}_N\bm{V}_N^{-1}(\bm{a}_N-\bm{A}_N)}{\sqrt{\left(\nabla g(\bm{A}_N)\right)^T \bm{V}_N\bm{V}_N\nabla g(\bm{A}_N)}}.
\end{align*}
So for our result to hold, it is sufficient that
\[\frac{\left(\nabla g(\bm{b}) - \nabla g(\bm{A}_N)\right)^T\bm{V}_N\bm{V}_N^{-1}(\bm{a}_N-\bm{A}_N)}{\sqrt{\left(\nabla g(\bm{A}_N)\right)^T \bm{V}_N\bm{V}_N\nabla g(\bm{A}_N)}} \xrightarrow[]{p} 0.\]
We have
\[\bm{V}_N^{-1}(\bm{a}_N - \bm{A}_N) \xrightarrow[]{d} \text{N}(\bm{0},\bm{I}_K)\]
so it is sufficient that 
\[\frac{\left(\nabla g(\bm{b}) - \nabla g(\bm{A}_N)\right)^T\bm{V}_N}{\sqrt{\left(\nabla g(\bm{A}_N)\right)^T \bm{V}_N\bm{V}_N\nabla g(\bm{A}_N)}} \xrightarrow[]{p} 0\]
(or $\bm{V}_N\left[\sqrt{\left(\nabla g(\bm{A}_N)\right)^T \bm{V}_N\bm{V}_N\nabla g(\bm{A}_N)}\right]^{-1}$ is bounded), which is a condition of Theorem 2.

\subsection{General case}
We now have the more general case for $\bm{A}_N, \bm{a}_N \in \mathbb{R}^K$, assuming that
\[\bm{a}_N - \bm{A}_N \xrightarrow[]{p} \bm{0}\]
and
\[\left(\frac{\bm{a}_{N[1]}-\bm{A}_{N[1]}}{\sqrt{\text{Var}(\bm{a}_{N[1]})}}, \dots, \frac{\bm{a}_{N[K]}-\bm{A}_{N[K]}}{\sqrt{\text{Var}(\bm{a}_{N[K]})}}\right)^T \xrightarrow[]{d} \text{N}(\bm{0},\bm{\Sigma}),\]
where $\bm{\Sigma}$ is the limit of the correlation matrix which we assume to be nonsingular.

We have
\[g(\bm{a}_N) = g(\bm{A}_N) + \left(\nabla g(\bm{b}_N)\right)^T(\bm{a}_N-\bm{A}_N)\]
for some $\bm{b}_N$ where $\bm{b}_{N[k]}$ is between $\bm{a}_{N[k]}$ and $\bm{A}_{N[k]}$ for each $k$, and where $\nabla g(\bm{b}_N)$ is the vector of partial derivatives evaluated at $\bm{b}_N$.

Then we have
\begin{align*}
g(\bm{a}_N) - g(\bm{A}_N) &= \left(\nabla g(\bm{b}_N)\right)^T (\bm{a}_N-\bm{A}_N)\\
 &= \left(\nabla g(\bm{b}_N)\right)^T \bm{V}_N\bm{V}_N^{-1}(\bm{a}_N-\bm{A}_N)\\
 &= \left(\nabla g(\bm{A}_N)\right)^T \bm{V}_N\bm{V}_N^{-1}(\bm{a}_N-\bm{A}_N)\\
 & \quad + \left(\nabla g(\bm{b}_N)-\nabla g(\bm{a}_N)\right)^T \bm{V}_N\bm{V}_N^{-1}(\bm{a}_N-\bm{A}_N).
 \end{align*}
 
 Then we can standardize to get
 \begin{align*}
\frac{g(\bm{a}_N) - g(\bm{A}_N)}{\sqrt{\left(\nabla g(\bm{A}_N)\right)^T \bm{V}_N\bm{\Sigma}\bm{V}_N\nabla g(\bm{A}_N)}}
 &= \frac{\left(\nabla g(\bm{A}_N)\right)^T \bm{V}_N}{\sqrt{\left(\nabla g(\bm{A}_N)\right)^T \bm{V}_N\bm{\Sigma}\bm{V}_N\nabla g(\bm{A}_N)}}\bm{V}_N^{-1}(\bm{a}_N-\bm{A}_N)\\
 & + \frac{ \left(\nabla g(\bm{b}_N)-\nabla g(\bm{A}_N)\right)^T \bm{V}_N}{\sqrt{\left(\nabla g(\bm{A}_N)\right)^T \bm{V}_N\bm{\Sigma}\bm{V}_N\nabla g(\bm{A}_N)}}\bm{V}_N^{-1}(\bm{a}_N-\bm{A}_N),
\end{align*}
where we have assumed that $(\nabla g(\bm{A}_N))^T \nabla g(\bm{A}_N) > 0$ for sufficiently large $N$.
 
Let's start with the first term of this final expression.
If $\bm{\Sigma}$ is nonsingular, we can write this term as 
\[ \frac{\left(\nabla g(\bm{A}_N)\right)^T \bm{V}_N\bm{\Sigma}^{1/2}}{\sqrt{\left(\nabla g(\bm{A}_N)\right)^T \bm{V}_N\bm{\Sigma}\bm{V}_N\nabla g(\bm{A}_N)}}\bm{\Sigma}^{-1/2}\bm{V}_N^{-1}(\bm{a}_N-\bm{A}_N).\]
We see that
\begin{align*}
\bm{\Sigma}^{-1/2}\bm{V}_N^{-1}(\bm{a}_N-\bm{A}_N) &= \bm{\Sigma}^{-1/2}\left(\frac{\bm{a}_{N[1]}-\bm{A}_{N[1]}}{\sqrt{\text{Var}(\bm{a}_{N[1]})}}, \dots, \frac{\bm{a}_{N[K]}-\bm{A}_{N[K]}}{\sqrt{\text{Var}(\bm{a}_{N[K]})}}\right)^T\\
& \xrightarrow[]{d} \text{N}(\bm{0},\bm{I}_K).
\end{align*}
Further, note that
\[\frac{\left(\nabla g(\bm{A}_N)\right)^T \bm{V}_N\bm{\Sigma}^{1/2}}{\sqrt{\left(\nabla g(\bm{A}_N)\right)^T \bm{V}_N\bm{\Sigma}\bm{V}_N\nabla g(\bm{A}_N)}}\]
is a unit vector that is constant for each $N$ but changes as $N \to \infty$.
So then we can use the same argument as in the previous section to show that
\[\frac{\left(\nabla g(\bm{A}_N)\right)^T \bm{V}_N\bm{\Sigma}^{1/2}}{\sqrt{\left(\nabla g(\bm{A}_N)\right)^T \bm{V}_N\bm{\Sigma}\bm{V}_N\nabla g(\bm{A}_N)}}\bm{\Sigma}^{-1/2}\bm{V}_N^{-1}(\bm{a}_N-\bm{A}_N)  \xrightarrow[]{d} \text{N}(0,1).\]

More generally, we can update the argument from Section~\ref{supp:simple} in the following ways: Take $\bm{X} \sim \text{N}(\bm{0},\bm{\Sigma})$. Then for every $N$, 
\[ \frac{\left(\nabla g(\bm{A}_N)\right)^T \bm{V}_N}{\sqrt{\left(\nabla g(\bm{A}_N)\right)^T \bm{V}_N\bm{\Sigma}\bm{V}_N\nabla g(\bm{A}_N)}}\bm{X} \sim N(0, 1).\]
Then note that because uniform continuity on a compact space implies that $\nabla g(\bm{A}_N)$ is bounded,
\[\frac{\left(\nabla g(\bm{A}_N)\right)^T \bm{V}_N}{\sqrt{\left(\nabla g(\bm{A}_N)\right)^T \bm{V}_N\bm{\Sigma}\bm{V}_N\nabla g(\bm{A}_N)}}\]
is bounded by the assumption in Theorem 2 that $\sqrt{\text{Var}(\bm{a}_{N[k]})}/\sqrt{\left(\nabla g(\bm{A}_N)\right)^T \bm{V}_N\bm{\Sigma}\bm{V}_N\nabla g(\bm{A}_N)} = O(1)$  for any $k$ such that $\nabla g(\bm{x})$ is not identically 0 for $\bm{x} \in \{R_N(\epsilon)\}$.

Now we need to show that the second term,
\[\frac{ \left(\nabla g(\bm{b}_N)-\nabla g(\bm{A}_N)\right)^T \bm{V}_N}{\sqrt{\left(\nabla g(\bm{A}_N)\right)^T \bm{V}_N\bm{\Sigma}\bm{V}_N\nabla g(\bm{A}_N)}}\bm{V}_N^{-1}(\bm{a}_N-\bm{A}_N),\]
goes to zero.

We have that 
\[\bm{V}_N^{-1}(\bm{a}_N-\bm{A}_N) \xrightarrow[]{d} \text{N}(\bm{0},\bm{\Sigma}),\]
so we need to show that the other factor goes to zero in probability.

We have by assumption that 
\[\frac{ \left(\nabla g(\bm{b}_N)-\nabla g(\bm{A}_N)\right)^T \bm{V}_N}{\sqrt{\left(\nabla g(\bm{A}_N)\right)^T \bm{V}_N\bm{\Sigma}\bm{V}_N\nabla g(\bm{A}_N)}}\xrightarrow[]{p} 0\]
In particular, recall that
\[\bm{b} - \bm{A}_N \xrightarrow[]{p} 0\]
and we have uniform continuity of the partial derivatives, so we can apply Theorem 2 of \cite{mann1943stochastic} and find that
\[\nabla g(\bm{b}_N) - \nabla g(\bm{A}_N) \xrightarrow[]{p} 0.\]
If we assume $\sqrt{\text{Var}(\bm{a}_{N[k]})}/\sqrt{\left(\nabla g(\bm{A}_N)\right)^T \bm{V}_N\bm{\Sigma}\bm{V}_N\nabla g(\bm{A}_N)} = O(1)$ for any $k$ such that $\nabla g(\bm{x})$ is not identically 0 for $\bm{x} \in \{R_N(\epsilon)\}$ we have 
\[\left(\nabla g(\bm{b}_N)-\nabla g(\bm{A}_N)\right)^T\frac{  \bm{V}_N}{\sqrt{\left(\nabla g(\bm{A}_N)\right)^T \bm{V}_N\bm{\Sigma}\bm{V}_N\nabla g(\bm{A}_N)}}\bm{V}_N^{-1}(\bm{a}_N-\bm{A}_N) \overset{p}{\to} 0\]
as desired.

Note that by Slutsky's theorem and the Continuous Mapping Theorem, we can replace the limiting value $\bm{\Sigma}$ with $\bm{\Sigma}_N$, the finite-population correlation matrix.

\section{Example of asymptotic degenerate distribution}
From Section 3.1, the asymptotic variance expression for a causal estimator that is a function of two treatment group means is \\$\nabla g_{[1]}^2\frac{v_{N,0}}{n_0} + \nabla g_{[2]}^2\frac{v_{N,1}}{n_1} - \frac{1}{N(N-1)}\sum_{i=1}^N\left[\nabla g_{[2]}Y_{N,i}(1) + \nabla g_{[1]}Y_{N,i}(0) - \left(\nabla g_{[2]}\bar{Y}_{N}(1) + \nabla g_{[1]}\bar{Y}_{N}(0)\right)\right]^2$.
Consider the case where $Y_i(1) = Y_i(0)$ for all $i$, for every $N$ and $p = 1/2$.
Then consider the estimand, $g(\bar{Y}(1), \bar{Y}(0)) = \bar{Y}(1) + \bar{Y}(0)$.
It is easy to show that $v_{N,0} = v_{N,1}$ and
\[\frac{1}{N-1}\sum_{i=1}^N\left[Y_{N,i}(1) + Y_{N,i}(0) - \left(\bar{Y}_{N}(1) +\bar{Y}_{N}(0)\right)\right]^2 = \frac{1}{N-1}\sum_{i=1}^N\left[2Y_{N,i}(0) - 2\bar{Y}_{N}(0) \right]^2 = 4v_{N,0} = 4v_{N,1}.\]

Thus
\begin{align*}
&\nabla g_{[1]}^2\frac{v_{N,0}}{n_0} + \nabla g_{[2]}^2\frac{v_{N,1}}{n_1} - \frac{1}{N(N-1)}\sum_{i=1}^N\left[\nabla g_{[2]}Y_{N,i}(1) + \nabla g_{[1]}Y_{N,i}(0) - \left(\nabla g_{[2]}\bar{Y}_{N}(1) + \nabla g_{[1]}\bar{Y}_{N}(0)\right)\right]^2\\
&=\frac{2v_{N,0}}{N} + \frac{2v_{N,1}}{N} - \frac{1}{N(N-1)}\sum_{i=1}^N\left(Y_{N,i}(1) + Y_{N,i}(0) - \left(\bar{Y}_{N}(1) + \bar{Y}_{N}(0)\right)\right]^2\\
&=\frac{2v_{N,0}}{N} + \frac{2v_{N,0}}{N} - \frac{4v_{N,0}}{N}\\
&=0.
\end{align*}

\end{appendices}

\end{document}